# A Survey of H-index, Stress, Tenure & Reference Management software use in Academia

Jose Berengueres
College of IT
UAE University
17551 Al Ain, A.D, U.A.E
jose@uaeu.ac.ae

Pavel Nesterov
Reaktor.fi
Dubai Mediacity
Dubai, U.A.E
pavel.nesterov@reaktor.fi

*Abstract*— We describe the findings of a survey that covered the topics of stress, citation tool use habits, subjective happiness, h-index, research topic and tenure among a sample of 2286 authors of arxiv.org. Ph.D. students report the lowest subjective happiness score among all faculty roles, while tenured faculty report the highest. Tenured faculty report the lowest levels of stress. Undergraduate and graduate students report the highest levels of stress. Non-tenured faculty report stress similar to postdocs. No association between citation management tool usage and h-index was found. The average age at tenure start is 34.9 years. In addition, no significant association between stress levels and the research topic was found.

## I. Introduction

In the past four decades (1980-2020), AI has progressively found its way to a plethora of industries under various names. From the *expert systems* that helped us reduce a foundry's waste output in the 1980s [1]; to the recommender systems popularized by the 2000 Netflix prize competition [2]. Today, AI, and particularly its successful rebrand as *Machine Learning,* (following the AI winter of 1980s [3]), seems to be powering productivity gains across society in the form of self-driving cars; better fraud detection and analytics in general [4, 5].

Given all this, it is not without irony, that the very same academics that helped create these AI/ ML tools are also the very last ones to enjoy their fruits at their own work. In fact, we are hard pressed to name a blockbuster case where AI helped make the job of a researcher easier. Some notable exceptions follow. In 2019, some timid evidence of how AI could assist in research was published in [6]. It describes one of first cases of NLP use to predict novel chemical compounds that did not exist three years earlier in the literature, the equivalent of an academic *Oracle*. This result was preceded a year earlier by a seminal paper [7] where the application of machine learning (predictive analytics) to research is foreseen. These, and subsequent papers, were initially circumscribed to the materials and chemical sciences. However, since 2020 we see a spillover to other fields such as life sciences [8-10, 38]. Another less conspicuous but also promising area primed to enhance the productivity of academics is reference management software. This field gained popularity with London based startup called Mendeley, founded in 2008. Mendeley and other reference tools (RefMan, EndNote, Zotero,) help to reduce paper writing time by managing the reference list, formatting and other overhead tasks. Reference management accounts to about **20%** of the total time spent writing a paper according to Citationsy.

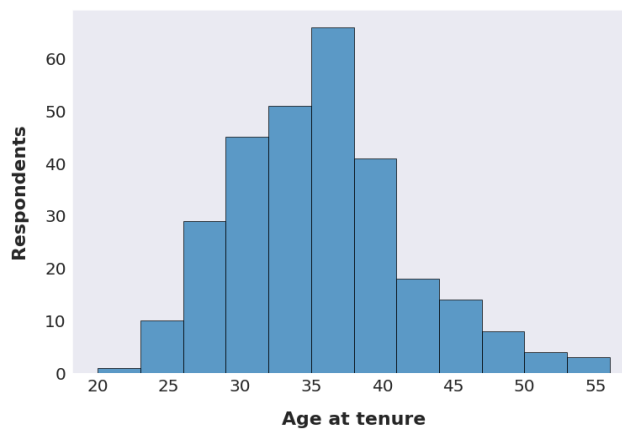

Fig. 1. The average age at which a respondent got tenured is 34.9, SD 6.5, sample size: 291 respondents that self-identified as faculty.

### A. Impact of AI

As shown by [6-8], it is not difficult to foresee how these productivity tools might improve with the acceleration of AI [11]. In the particular case of reference management and discovery, a natural pathway is recommendation on what papers to read next. Interestingly, AI-driven recommendation systems similar to the Facebook feed are already live in the likes of Google Scholar™ and Research Gate™. Given the controversial impact to these AI-driven feeds on mental health [29-37], it is worth considering not only their impact on research but also on the well-being of the person. Surprisingly, very little is known about the impact of these tools on academics. Are academics that use reference management tools more productive? Are they happier? A literature search yields very few results. [12] evaluated **ethical** conflicts of interest found in Google Scholar algorithms; [13] compared the existing tools; [14] investigated how ResearchGate metrics can be used to infer the performance of a given researcher. However, none considers well-being. Following, we analyze a survey conducted by reference tool Citationsy Ltd. in collaboration with UAE University during the summer of 2020. We hope the findings will help inform the impact research tools on academics while considering as well well-being aspects.

## II. Methods

### A. Data Collection

Data of the survey was collected from 2020-07-09 to 2020-09- 28. The data was generated by emailing authors who had authored or co-authored at least one paper in a popular preprint server service in the past 3 years (Arxiv)[15]. In total, **2286** responses were collected. The average age of the respondents is 36.4 years, SD=11.42, sample size 200. The survey consisted of 10 questions. Table 1 shows the questions of the survey. Questions 4 to 7, (shaded), correspond the four standard questions used to compute the *subjective happiness* scale by [16]. Table 2 shows data about the papers of the respondents. A surprising fact from an initial exploration of the data is that 12% of the respondents (authors) identify themselves as employees rather than academics. Regarding faculty, 293 out of 386 respondents report to be tenured (76%). The average age at which tenure was obtained is 34.9 years old, SD = 6.5 years, sample size 291 (See Fig. 1).

*B. Data Cleaning*

Following is a highlight of the most relevant data challenges encountered.

(i) In question 9, due to the proactive nature of the participants, many decided to specify their job role beyond the faculty, Ph.D., postdoc and student prompts offered as defaults. This resulted in responses with a remarkable granularity that reflects the complexity of academic world. All these unexpected roles (doctor, retired doctor, retired faculty, C.E.Os., professor emeritus …), have been mapped to five main roles (See 'roles' in Table 2). The mapping of roles can be found in the file roles.txt in the Appendix.

(ii) In the same vein, for the question *at what age did you get tenure* we received respondent feed-back that not all countries have equivalent tenure track systems. However, we do not consider different tenure systems.

(iii) Data **attrition**. Due to technical and anonymization hurdles, from the 2286 responses, only 1016 were matched to a research subject. Hence, some statistics related to disciplines are calculated on sample size of 1016 or lower depending of the completeness of responses or subset considered.

(iv) The topic of an authored paper was assigned as the first category out of the maximum of four category options available on Arxiv.

(v) Outliers such as very high self-reported h-index (>1000), self-evident typos and so on, are excluded from the calculations (~12 cases).

*C. Ethics review board & disclosures*

This study received ethics approval by the ethics board at the Research Office at UAE university (id: ERS_2020_6162). Disclosures and conflicts of interest are as follows: The first author is investor in the company that conducted the survey (Citationsy Ltd).

*D. Satistical methods*

As we are analyzing multiple differences in means between disciplines the increased chance of spurious correlations must be accounted for as explained in depth in [17]. One way to do so is by using the Bonferroni correction method. We used the Python library *statsmodels* (v0.13.0.dev0 +36). A detailed statistical analysis is available in the Appendix.

TABLE I. SURVEY QUESTIONS

| Question | Scale | Mean SD [b] |
|---|---|---|
| How stressed are you because of work? [1 not at all - 7 a great deal] | 7-Likert | **4.20** 1.64 |
| What is your h-index? | Numeric | 12.6 44.0 |
| At what age did you get tenure? | Numeric | 33.4 - |
| In general, I consider myself... [1 less happy – 7 more happy] | 7-Likert [a] | 5.0 1.31 |
| Compared with most of my peers, I consider myself: [1 less happy – 7 more happy] | 7-Likert [a] | 4.8 1.34 |
| Some people are generally very happy. They enjoy life regardless of what is going on, getting the most out of everything. To what extent does this characterization describe you? [1 less happy – 7 a great deal] | 7-Likert [a] | 4.32 1.62 |
| Some people are generally not very happy. Although they are not depressed, they never seem as happy as they might be. To what extent does this characterization describe you? [1 less happy – 7 a great deal] | 7-Likert [a] | 3.14 1.65 |
| Do you use any reference management tool? [1 no, never – 5 for every paper] | 5-Likert | 2.41 1.58 |
| Would you like to help in a next survey ? | Y/N | 55% Yes |
| Your role is... (faculty, postdoc, student… | Free text | - |
| Any other comments, suggestions, or questions you'd add... | Free text | 9.8% wrote a comment |

a. These questions correspond to the standard subjective happiness scale survey by [16]
b. Where otherwise stated the first figure is mean and the second is standard deviation

TABLE II. OVERVIEW OF RESPONSES BY PAPER

| Item | Most frequent items by count (%) |
|---|---|
| **Categories** (17 different categories [b]) | Computer Science 426<br>Mathematics 192<br>Physics 77<br>Astrophysics 66<br>Condensed Matter 60 |
| **Title** (3 943 different keywords [a]) | learning 60 (5%) [c]<br>neural 42<br>multi 31<br>model 29<br>data 27<br>quantum 26<br>deep 25 (2.4%) |
| **e-mail** (83 different domains) [a] | .edu 219 (21%)<br>.com 88 ( 8%)<br>.de 72<br>.uk 53<br>.in 49<br>.fr 48<br>.ca 46 |
| **Roles** [b] (25 different roles mapped to 5) | faculty 390 (40%)<br>Ph.D. student 171 (17%)<br>researcher/postdoc. 171<br>employee 124 (12%)<br>student 109 (11%) |

a. Not shared for anonymity purposes.

b. Sample considered 975.
c. Percent of paper with this keyword in the title , sample 1016.

## III. RESULTS

No significant correlation is found between h-index, stress, subjective happiness and reference management software use frequency. As expected, a moderate correlation of **-.45** is found between *stress* and *subjective happiness*. Following we explain the findings by role, subject and email domain of authors.

### A. By Role

Globally, respondents scored on average a subjective happiness of **4.76**. This figure is similar to scores reported by students [18] (4.8 and 4.9), but **lower** than scores reported by working professionals (health care, 5.2) [19]. Among academic roles, tenured faculty reported the highest subjective happiness at **4.96**; Ph.D. students reported the lowest score of all academic groups at **4.5**. Tenured faculty reported the lowest stress at work while the rest of the groups reported similar stress. Students (graduate and undergraduate) reported the most stress attributed to work. Table 3 shows a detailed breakdown. Faculty is the role least likely to use a reference management software. Fig. 2 is a density plot of stress that visualizes why tenured faculty shows a lower stress than other groups. Among tenured faculty two subgroups exist. One reported stress similar to non-tenured faculty (**5.0**), while a second group reported stress near **2.0** (See 'faculty tenured' label in Fig. 2). Fig. 3 shows a scatter plot of roles. We note that in terms of stress and subjective happiness, (i) postdocs, (ii) researchers and (iii) non-tenured faculty are clustered close together (the label 'non-tenured' includes postdocs, researchers and non-tenured faculty). In addition, in terms of stress, (Ph.D. students, undergraduates, graduates, employees, non-tenured faculty and postdocs), show similar stress levels but varying degrees of subjective happiness.

### B. By research subject

Table 4 shows a breakdown by the research subject of the paper associated to the respondent. Only the top 10 are shown for brevity. This table is for illustration purposes and the aggregates are not significative. Table 5 shows a summary of the Bonferroni correlation analysis. It yields no significant correlation between research subject and stress. In other words, there are no happier research disciplines than others.

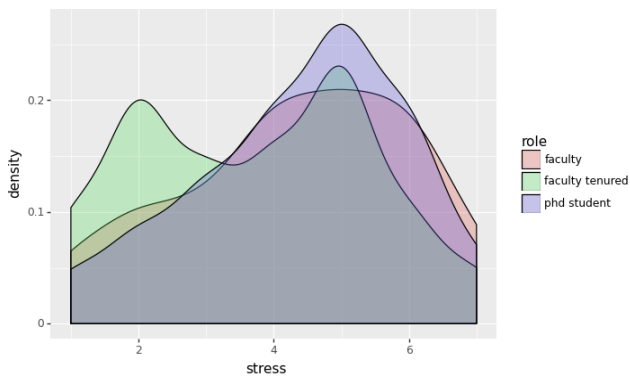

Fig. 2. Non-tenured faculty (label 'faculty'), and PhD students report similar stress distributions. Tenured faculty can be divided in two subgroups one with very low stress arround 2.0 and another with a stress similar to the other non-tenured academics (See table 3)

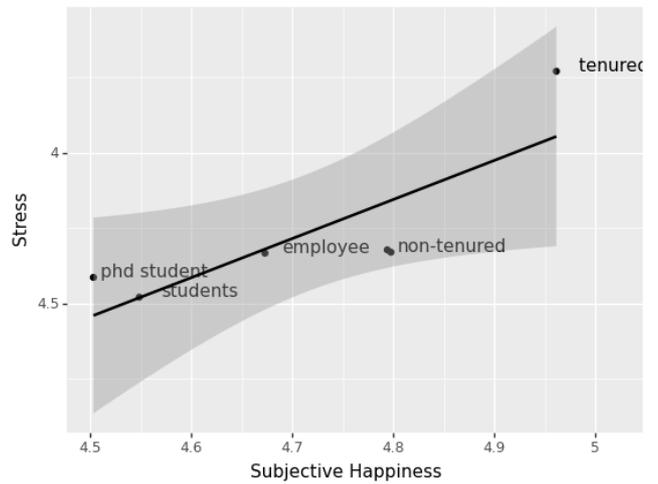

Fig. 3. Ph.D. students report the lowest subjective happiness among all academic roles. Tenured faculty reports the lowest stress among all roles. Note: non-tenured label includes non-tenured faculty and postdocs for label clarity. Tenured label represents tenured faculty. 95% CI shown in shade. (data from table 3)

TABLE III. STRESS, PERFROMANCE AND USAGE BY ROLE

|  | subjective happiness | stress work[a] | Ref. tools usage[b] | H-index | Count |
|---|---|---|---|---|---|
| Faculty **tenured** | 4.96 | **3.73** | 2.1 | 20.8 | 348 |
| Faculty non-tenured | 4.80 | 4.33 | 2.5 | 11.3 | 162 |
| Employee | 4.79 | 4.32 | 2.4 | 7.6 | 149 |
| Researcher, postdoc | 4.67 | 4.33 | 2.6 | 7.2 | 107 |
| Ph.D. student | **4.50** | 4.41 | 2.5 | 3.0 | 100 |
| Student grad, undergrad | 4.55 | 4.48 | 2.4 | 2.6 | 99 |

a. 7 point Likert scale
b. 5 points Likert scale]

TABLE IV. STRESS, PERFROMANCE, RESERCH SUBJECT

| Topic | Average stress | Average h-index | Count |
|---|---|---|---|
| Electrical Engineering and Systems Science | 4.4 | 8.3 | 40 |
| Computer Science | 4.4 | 9.1 | 407 |
| Statistics | 4.4 | 7.8 | 33 |
| Quantum Physics | 4.3 | 11.4 | 17 |
| Physics | 4.0 | 14.9 | 71 |
| High Energy Physics | 4.0 | 26.9 | 29 |
| Condensed Matter | 3.9 | 15.7 | 58 |
| Mathematics | 3.9 | 8.5 | 179 |
| General Relativity and Quantum Cosmology | 3.8 | 27.6 | 19 |

TABLE V. TOP CORRELATIONS WITH CORRECTED P-VALUES

| subj_a | subj_b | Mean a | Mean b | p-value | p-value Corrected [a] |
|---|---|---|---|---|---|
| Astrophysics | Computer Science | 3.74 | 4.41 | 0.002 | **0.28** |
| Computer Science | Mathematical Physics | 4.41 | 2.93 | 0.004 | **0.56** |
| Computer Science | Mathematics | 4.41 | 3.87 | .0002 | **0.03** |
| Electrical Engineering | Mathematical Physics | 4.42 | 2.93 | 0.005 | **0.81** |

[a.] Corrected Bonferroni

### C. By e-mail domain

Fig. 4 shows a scatter plot of stress vs. subjective happiness. Cultural biases as well as different country-tenure systems might explain the difference between e-mail domains. The moderate association between subjective happiness and stress seems to hold at country level as well. Academics with e-mail addresses from Germany, U.K. and Israel, report to be the least happy and report the most stress. Iran, Japan and India are on the opposite side of the spectrum.

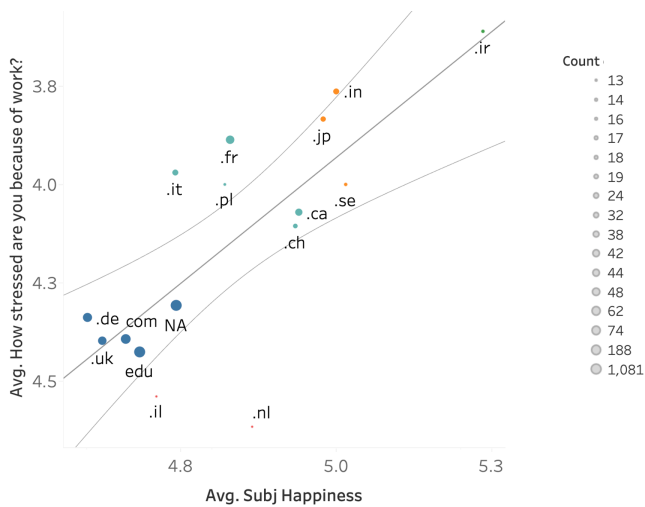

Fig. 4. A scatter plot of stress vs. subjective happiness. All roles considered. Only items with N>13 are shown. NA label: e-mail not available (any domain).

## IV. DISCUSSION

### A. H-index

No significant association between h-index and how often an author used reference management software tools was found. This somehow counterintuitive fact could be partly explained by the fact that high h-index faculty tend to be older and hypothetically less adept to use these tools as compared to younger groups such as milenials or Gen-Z students. In addition, h-index association with reference management software, if exists, might only show up years later in one's career due to the multi-year dynamics of h-index formula [25, 26].

### B. Research subject

The topic of the respondent's research was **not** found to be associated with stress. This came as a surprise, as we expected fields with high popularity and high **employment prospects** to be associated with lower stress. Nevertheless, results here agree with results found elsewhere in the literature that report that an individual's well-being depends on factors such as for example the environment and attitudes towards life [20, 21], seniority [27], and beliefs [28] among many other factors. The results here also support the view that subjective happiness is not associated with the research topic (as expected).

### C. Job security

Half of the tenured faculty report a stress near 2.0, more than one SD lower than the average of 4.2 (SD = 1.64). Regarding subjective happiness, tenured and non-tenured faculty show a similar distribution. This is in spite that tenured faculty tend to be older and that some surveys show that happiness declines with age [23, 24]. This seems to confirm the robustness of the subjective happiness scale [16].

### D. Limitations

No gender or age was considered explicitly. Age is considered implicitly in the h-index and with the age at tenure which only applies to faculty roles. Each paper can be classified by up to four subjects. However, only the first subject is used to associate an author's responses to their research subject. This might not be accurate for interdisciplinary authors. Users of preprint servers might not be representative of academia in general. This survey might have other unaccounted biases such selection bias, country specific bias.

## V. CONCLUSIONS

The main conclusions are four: (i) the stress is correlated -.45 with subjective happiness index. (ii) Tenure is a factor in stress. (iii) No significant correlation was found between reference software usage and h-index. (iv) No association between topic of the paper and author's stress.

ACKNOWLEDGMENT

We would like to thank Cenk Dominic Özbakır for technical support and to the participants in the survey. In particular the ones whose feedback helped improve the design of the survey.

APPENDIX

Anonymized dataset and python notebook are available at https://www.kaggle.com/harriken/mentalhealth-academics/

REFERENCES

[1] Chandrasekaran, B., 1986. Generic tasks in knowledge-based reasoning: High-level building blocks for expert system design. *IEEE expert*, *1*(3), pp.23-30.
[2] Bennett, J. and Lanning, S., 2007, August. The netflix prize. In *Proceedings of KDD cup and workshop* (Vol. 2007, p. 35).
[3] Hendler, J., 2008. Avoiding another AI winter. *IEEE Intelligent Systems*, (2), pp.2-4.
[4] E. Siegel, *Predictive analytics*. Hoboken: John Wiley & Sons, 2016.
[5] T. Davenport, Competing on Analytics: Updated, with a New Introduction. Harvard Business Press, 2017.
[6] V. Tshitoyan *et al.*, "Unsupervised word embeddings capture latent knowledge from materials science literature," *Nature*, vol. 571, no. 7763, pp. 95–98, 2019, doi: 10.1038/s41586-019-1335-8.

[7] Butler, K. T., Davies, D. W., Cartwright, H., Isayev, O. & Walsh, A. Machine learning for molecular and materials science. Nature 559, 547–555 (2018).

[8] Venkatakrishnan, A.J., Puranik, A., Anand, A., Zemmour, D., Yao, X., Wu, X., Chilaka, R., Murakowski, D.K., Standish, K., Raghunathan, B. and Wagner, T., 2020. Knowledge synthesis from 100 million biomedical documents augments the deep expression profiling of coronavirus receptors. *arXiv preprint arXiv:2003.12773*.

[9] Stokes, J.M., Yang, K., Swanson, K., Jin, W., Cubillos-Ruiz, A., Donghia, N.M., MacNair, C.R., French, S., Carfrae, L.A., Bloom-Ackerman, Z. and Tran, V.M., 2020. A deep learning approach to antibiotic discovery. *Cell*, 180(4), pp.688-702.

[10] Goecks, J., Jalili, V., Heiser, L.M. and Gray, J.W., 2020. How machine learning will transform biomedicine. *Cell*, 181(1), pp.92-101.

[11] Tang, X., Li, X., Ding, Y., Song, M. and Bu, Y., 2020. The pace of artificial intelligence innovations: Speed, talent, and trial-and-error. *Journal of Informetrics*, 14(4), p.101094.

[12] Jacsó, P., 2005. Google Scholar: the pros and the cons. Online information review.

[13] Butros, A. and Taylor, S., 2010, October. Managing information: evaluating and selecting citation management software, a look at EndNote, RefWorks, Mendeley and Zotero. In Netting knowledge: two hemispheres/one world: proceedings of the 36th IAMSLIC Annual Conference (pp. 17-21). USA: IAMSLIC.

[14] Yu, M.C., Wu, Y.C.J., Alhalabi, W., Kao, H.Y. and Wu, W.H., 2016. ResearchGate: An effective altmetric indicator for active researchers?. *Computers in human behavior*, 55, pp.1001-1006.

[15] E. Gibney, "Open journals that piggyback on arXiv gather momentum," Nature, vol. 530, no. 7588, pp. 117–118, 2016, doi: 10.1038/nature.2015.19102.

[16] 2b S. Lyubomirsky and H. S. Lepper, "A measure of subjective happiness: Preliminary reliability and construct validation," vol. 46, no. 2, pp. 137–155, 1999, doi: [10.1023/a:1006824100041](10.1023/a:1006824100041).

[17] Armstrong, R.A., 2014. When to use the B onferroni correction. *Ophthalmic and Physiological Optics*, 34(5), pp.502-508.

[18] Yue, X.D., Leung, C.L. and Hiranandani, N.A., 2016. Adult playfulness, humor styles, and subjective happiness. *Psychological Reports*, 119(3), pp.630-640.

[19] Ruiz‐Aranda, D., Extremera, N. and Pineda-Galan, C., 2014. Emotional intelligence, life satisfaction and subjective happiness in female student health professionals: the mediating effect of perceived stress. *Journal of psychiatric and mental health nursing*, 21(2), pp.106-113.

[20] Krieger, L.S., 2004. The inseparability of professionalism and personal satisfaction: Perspectives on values, integrity and happiness. *Clinical L. Rev.*, 11, p.425.

[21] Walsh, L.C., Boehm, J.K. and Lyubomirsky, S., 2018. Is happiness a consequence or cause of career success?. *LSE Business Review*.

[22] Joo, B.K. and Lee, I., 2017, August. Workplace happiness: Work engagement, career satisfaction, and subjective well-being. In *Evidence-based HRM: A Global Forum for Empirical Scholarship*. Emerald Publishing Limited.

[23] Vera-Villarroel, P., Celis-Atenas, K., Pavez, P., Lillo, S., Bello, F., Díaz, N. and López, W., 2012. Money, age and happiness: association of subjective wellbeing with socio-demographic variables. *Revista latinoamericana de Psicología*, 44(2), pp.155-163.

[24] Quezada, L., Landero, R. and Gonzalez, T., 2016. A validity and reliability study of the subjective happiness scale in Mexico. *The Journal of happiness & Well-Being*, 4(1), pp.90-100.

[25] Egghe, L. and Rousseau, R., 2008. An h-index weighted by citation impact. Information Processing & Management, 44(2), pp.770-780.

[26] Costas, R. and Franssen, T., 2018. Reflections around 'the cautionary use' of the h-index: Response to Teixeira da Silva and Dobránszki. *Scientometrics*, 115(2), pp.1125-1130.

[27] Bebenroth, R. and Berengueres, J.O., 2020. New hires' job satisfaction time trajectory. *International Journal of Human Resources Development and Management*, 20(1), pp.61-74.

[28] Caprara, G.V., Steca, P., Gerbino, M., Paciello, M. and Vecchio, G.M., 2006. Looking for adolescents' well-being: Self-efficacy beliefs as determinants of positive thinking and happiness. *Epidemiology and Psychiatric Sciences*, 15(1), pp.30-43.

[29] *The Social Dilema*. Larissa Rhodes.

[30] Light, B. and McGrath, K., 2010. Ethics and social networking sites: a disclosive analysis of Facebook. *Information Technology & People*.

[31] Rosenquist, J.N., Fowler, J.H. and Christakis, N.A., 2011. Social network determinants of depression. *Molecular psychiatry*, 16(3), pp.273-281.

[32] Olsson, G.I., Nordström, M.L., Arinell, H. and Von Knorring, A.L., 1999. Adolescent depression: social network and family climate—a case-control study. *Journal of child psychology and psychiatry*, 40(2), pp.227-237.

[33] Yoon, S., Kleinman, M., Mertz, J. and Brannick, M., 2019. Is social network site usage related to depression? A meta-analysis of Facebook–depression relations. *Journal of affective disorders*, 248, pp.65-72.

[34] Li, Y., 2019. Upward social comparison and depression in social network settings. *Internet Research*.

[35] Domènech-Abella, J., Mundó, J., Haro, J.M. and Rubio-Valera, M., 2019. Anxiety, depression, loneliness and social network in the elderly: Longitudinal associations from The Irish Longitudinal Study on Ageing (TILDA). *Journal of affective disorders*, 246, pp.82-88.

[36] Santini, Z.I., Jose, P.E., Cornwell, E.Y., Koyanagi, A., Nielsen, L., Hinrichsen, C., Meilstrup, C., Madsen, K.R. and Koushede, V., 2020. Social disconnectedness, perceived isolation, and symptoms of depression and anxiety among older Americans (NSHAP): a longitudinal mediation analysis. *The Lancet Public Health*, 5(1), pp.e62-e70

[37] Carbonell, X. and Panova, T., 2017. A critical consideration of social networking sites' addiction potential. *Addiction Research & Theory*, 25(1), pp.48-57.

[38] McCreery, C.H., Katariya, N., Kannan, A., Chablani, M. and Amatriain, X., 2020, August. Effective Transfer Learning for Identifying Similar Questions: Matching User Questions to COVID-19 FAQs. In *Proceedings of the 26th ACM SIGKDD International Conference on Knowledge Discovery & Data Mining* (pp. 3458-3465).